\documentstyle[aps,prl,graphicx,multicol]{revtex}

\def\vec#1{{\bf #1}}

\def\cale{\mbox{\boldmath$\cal E$}}
\def\calb{\mbox{$\cal B$}}
\def\state#1{|#1 \mbox{ Fermions} \rangle}
\def\beps{\mbox{\boldmath$\epsilon$}}
\def\half{{{1}\over{2}}}

\begin{document}

\title{Coexistence of
Composite-Bosons and Composite-Fermions in
  $\nu=\half + \half$ Quantum Hall Bilayers}

\author{Steven H. Simon${}^a$, E. H. Rezayi${}^{b}$,  and Milica Milovanovic${}^c$}

\address{${}^a$ Lucent Technologies, Bell Labs, Murray Hill, NJ, 07974
\\${}^b$Department of Physics, California State University,
 Los Angeles California 90032
\\${}^c$  Institute of Physics, P.O. Box 68, 11080, Belgrade,
Yugoslavia}

\date{\today}
\maketitle

\begin{abstract}
In bilayer quantum Hall systems at filling fractions near
$\nu=1/2+1/2$, as the spacing $d$ between the layers is
continuously decreased, intra-layer correlations must be replaced
by inter-layer correlations, and the composite fermion (CF) Fermi
seas at large $d$ must eventually be replaced by a composite boson
(CB) condensate or ``111 state" at small $d$.   We propose a
scenario where CBs and CFs coexist in two interpenetrating fluids
in the transition. Trial wavefunctions describing these mixed
CB-CF states compare very favorably with exact diagonalization
results. A Chern-Simons transport theory is constructed that is
compatible with experiment.
\end{abstract}

\begin{multicols}{2}

Bilayer quantum Hall systems show a remarkable variety of
phenomena\cite{Pinczuk}.  Perhaps the most studied case is when
the electron density in each of the two layers is such that $\nu=n
\phi_0/B=\half$, where $n$ is density, $\phi_0 = 2 \pi \hbar c/e$
is the flux quantum and $B$ is the magnetic field
perpendicular to the sample.   At this filling fraction, it is
known that at least two types of states can occur depending on the
spacing $d$ between the layers. For large $d$ the two layers must
be essentially independent $\nu=\half$ states, which are thought
to be well described as compressible composite fermion (CF) Fermi
seas with strong intralayer correlations and no interlayer
correlations\cite{Olle}. For small enough values of $d$ one should
have an interlayer coherent ``111 state'' which can be described
as a composite boson (CB) condensate with strong interlayer
correlations and intralayer correlations which are weaker than
that of the CF Fermi sea\cite{Pinczuk}. The nature of the
transition between CFs and CBs is the focus of this paper.
(Throughout this paper we will assume zero interlayer tunnelling,
assume the spins are fully polarized, and set the in-plane field
to zero.)

Previous numerical work\cite{Girvin} suggested that the transition
between the CF Fermi sea and the CB 111 state may be first order.
(A number of more exotic mechanisms have also been
proposed\cite{Exotic}).  However, experiments clearly show that
interlayer correlations and coherence turn on somewhat
continuously as $d/\ell_B$ is
reduced\cite{Eisenstein,Kellogg1,Kellogg2} ($\ell_B = (\hbar e/B
c)^{1/2}$ is the magnetic length).    Based on  a picture of a
first order transition and percolating puddles of one phase within
the other, Ref. \cite{SternHalperin} predicts the drag resistivity
tensor $\rho^D$ should roughly obey the semicircle relation $
(\rho^D_{xx})^2  + (\rho^D_{xy} +  \pi \hbar /e^2)^2 = ( \pi \hbar
/e^2)^2 $ which agrees reasonably well with
experiment\cite{Kellogg2}.   On the other hand, the first order
transition model has trouble accounting for the strong interlayer
correlations that appear to occur even deep into the putative
Fermi liquid state\cite{Kellogg2}. We are thus motivated to look
for a more continuous transition from the CF Fermi liquid to the
CB 111 state. The picture we have in mind is a family of states
interpolating between these endpoints where each state is
specified by the number of CFs and the number of CBs with the
total number of CBs plus CFs remaining fixed.   A first order
transition could also be described as a mixture of CBs and CFs but
with phase separation of the two fluids. Here we instead consider
states where the CF and CB fluids interpenetrate. Since the
wavefunction must be fully antisymmetric in terms of the original
electrons, one might think of the electron having some CF
character and some CB character.  This interpolation between CF
and CB is advantageous since it allows us (by varying the ratio of
CFs to CBs) to construct states with varying degrees of inter-
versus intra-layer correlations. We find that such mixed CF-CB
states agree well with exact diagonalizations. Further, we find
that a Chern-Simons version of our mixed Bose-Fermi theory is
consistent with experimental observation, and in particular also
predicts the above mentioned semicircle relation.

{\em CF Fermi Sea and 111 State:}  Near $\nu=\half$ in a single
layer, the Jain CF\cite{Olle} picture  is given by attaching 2
zeros of the wavefunction to each particle. Thus we write the
electron wavefunction as $
 \Psi =  {\cal P} \left[
\Phi_f[z_1,\overline{z}_1, \ldots, z_N, \overline{z}_N] \prod_{i <
j} (z_i - z_j)^{2} \right] $ where ${\cal P}$ represents
projection onto the lowest Landau level, where for convenience
here and elsewhere the gaussian factors $\exp[-\sum_i |z_i|^2/(4
\ell_B^2)]$ will not be written explicitly, and $z_i = x_i + i
y_i$ is the complex representation of the position. In the above
equation $\Phi_f$ represents the fermionic wavefunction of the CFs
in the effective magnetic field $
    {\calb} = B - 2 n \phi_0 $
where $n$ is the density.  Generally, we will assume the CFs are
weakly interacting so that the wavefunction $\Phi_f$ can be
written as a single Slater determinant appropriate for
noninteracting fermions in the effective field $\calb$.  At
$\nu=1/2$, $\cal B$ is zero and $\Phi_f$ represents a filled Fermi
sea wavefunction.

In Chern-Simons fermion theory\cite{HLR,Olle}, each electron is
exactly transformed into a fermion bound to 2 flux quanta.  At
mean field level, the fermions see magnetic field $
    {\calb} = B - 2 n \phi_0 $ which is zero at $\nu=\half$.
The effective (mean) electric field seen by a fermion is given
analogously by $\cale = \vec E - 2 \beps \vec J$ where $\vec E$ is
the actual electric field, $\vec J$ is the fermion current (which
is equal to the electrical current) and $\beps = (2 \pi \hbar/e^2)
\tau$ where $\tau = i \sigma_y$ is the 2 by 2 antisymmetric unit
tensor (here $\sigma_y$ is the Pauli matrix).
Defining a transport equation for the weakly interacting fermions
$\cale = \rho_f \vec J$ (where $\rho_f$ is approximated as simple
Drude or Boltzmann transport for fermions in zero magnetic field),
we obtain the RPA expression for the electrical resistivity $\rho
= \rho_f + 2 \beps$.

We now turn to the double layer systems and focus on filling
fraction $\nu_1 = \nu_2= \frac{1}{2}$.  If the two layers are very
far apart, then we should have two independent $\nu=\frac{1}{2}$
systems. Thus, we would have a simple CF liquid state in each
layer with the total wavefunction being just a product of the
wavefunctions for each of the two layers. It should be noted that
such a state completely neglects the correlation effects of the
interlayer interaction.

On the other hand, if the two layers are brought very close
together, the intra- and inter-layer interactions will be roughly
the same stength. In this case, it is known that the system will
instead be described by the so-called 111 state.  The wavefunction
for this state is written as $
  \Psi = \prod_{i < j}  (z_i - z_j)
  \prod_{i < j}(w_i - w_j)
  \prod_{i,j}  (z_i - w_j)
$
%
%
%
where the $z_i$'s represent the electron coordinates in the first
layer and the $w_i$'s represent electron coordinates in the second
layer.   Here each electron is bound to a single zero of the
wavefunction within its own layer as well as being bound to a
single zero of the wavefunction in the opposite layer. Of course
by Fermi statistics, each electron must be bound to at least one
zero within its own layer so the only additional binding here is
interlayer. Thus, when $d$ is small, the electron binds a zero in
the opposite layer (to form a CB inter-layer dipole) whereas when
$d$ is large the electron minimizes its energy by binding to a
zero within its own layer (forming a CF dipole\cite{Olle}).

One can also write a Chern-Simons boson theory for the 111 state.
Here, each electron is exactly modeled as a boson bound to 1 flux
quantum where the bosons see the Chern-Simons flux from both
layers. Thus, the effective magnetic (mean) field seen by a boson
in layer $\alpha$ is given by $
  {\cal B}^\alpha = B - \phi_0 ( n^1 + n^2)
$ with $n^i$ being the density in layer $\alpha$.   Here, at total
filling fraction of $\nu_1 + \nu_2 = 1$, the bosons in each layer
see an effective field of zero and can condense to form a
superfluid (or quantum Hall) state.  The effective electric field
seen by the bosons in layer $\alpha$ is similarly given by
$\cale^\alpha = \vec E^\alpha -  \beps (\vec J^1 + \vec J^2)$
where $\vec E^\alpha$ is the actual electric field in layer
$\alpha$. This can be supplemented by a transport equation for the
bosons in zero magnetic field $\cale^\alpha = \rho_b^\alpha \vec
J^\alpha$.   If the bosons are indeed condensed, then we can set
$\rho^\alpha_b=0$. This results in the perfect Hall drag
indicative of the 111 state where $\vec E^1 = \vec E^2 = \beps
(\vec J^1 + \vec J^2)$.

{\em{Transition Wavefunctions:}}  At intermediate $d$ we need to
ask what the energetic price is for binding within the layer (to
form a CF) versus out of the layer (to form a CB). The important
thing to note is that (as we might expect for fermions) each
fermion put into the Fermi sea costs successively more energy
(each having a higher wavevector than the last). Thus, it might be
advantageous for some of the fermions at the top of the Fermi sea
to become unbound from their zeros within the layer and bind to
the other layer --- falling into the boson condensate.  We imagine
having some number $N_f^{\alpha}$ of electrons in layer $\alpha$
that act like CFs (filling a Fermi sea) and $N_b^{\alpha}$ that
act like CBs (which can condense). Of course we should have
$N^\alpha_f + N^\alpha_b = N^\alpha$ the total number of electrons
in layer $\alpha$. We can write down mixed Fermi-Bose
wavefunctions for double layer systems as follows:
\end{multicols}
\begin{eqnarray}  \label{eq:transitionwf}
 \Psi &=& {\cal A \, P}  \left[ \rule{0pt}{25pt} \, \Phi_{f}^1[z_1,\overline{z}_1, \ldots, z_{N_f^1},
 \overline{z}_{N_f^1}]
  \, \Phi_f^2[w_1, \overline{w}_1, \ldots, w_{N_f^2}, \overline{w}_{N_f^2}]\prod_{i < j \le N_f^1}
  (z_i - z_j)^2
 \prod_{i < j \le N_f^2} (w_i - w_j)^2 \,
 \times \right. \\  \nonumber
& &  \left.
\prod_{j <
i; N_f^1 < i} (z_i - z_j)   \prod_{j < i; N_f^2 < i} (w_i - w_j)
   \prod_{i, j; N_f^1 < i} (z_i - w_j) \prod_{i,j; N_f^2 < i; j \leq N_f^1} (w_i - z_j)
  \right]_.
\end{eqnarray}
\begin{multicols}{2}
We have chosen to order the particles so that particles $i = 1,
\ldots N_f^\alpha$ are fermions and $i=N_f^\alpha + 1, \ldots,
N^\alpha$ are bosons. The antisymmetrization  operator ${\cal A}$
antisymmetrizes only over particle coordinates within each layer,
and $\cal P$  is the lowest Landau level projection.    Here,
$\Phi_f^{\alpha}$ is the CF wavefunction of $N^{\alpha}_f$
fermions in layer $\alpha$.  The first line of the wavefunction is
thus the fermionic part, including the CF wavefunctions and also
the Jastrow factors.  Here the Jastrow factors bind two zeros to
each fermion within a layer in the sense that the wavefunction
vanishes as $z^2$ as two fermions in the same layer approach each
other (before antisymmetrization and projection). The second line
of the wavefunction binds zeros to each boson such that the
wavefunction vanishes as $z$ as any particle (boson or fermion)
approaches that boson from either layer.  (The bosons are assumed
condensed so the explicit boson wavefunction is unity).

It is perhaps easier to describe these Jastrow factors in terms of
an effective Chern-Simons description.   Within such a
description, we write expressions for the effective magnetic field
${\calb}^\alpha$ seen by either species in layer $\alpha$ as
\begin{eqnarray} \label{eq:Beff}
  {\cal B}^{\alpha}_f &=& B - 2 \phi_0 n^\alpha  - \phi_0 (n_b^1 + n_b^2) \\
  {\cal B}^{\alpha}_b &=& B -  \phi_0 (n_b^1 + n_b^2 + n_f^1
  + n_f^2)_. \label{eq:Beff2}
\end{eqnarray}
We note, however, that unlike in the prior 111 case and Fermi
liquid case, the transformation to a Chern-Simons theory  here is
not exact (even before making a mean field approximation) because
of complications associated with arbitrarily choosing which
particles are bosons and which are fermions.  It seems plausible
that as a low energy description, to the extent that zeros are
bound tightly to electrons, such a mixed Bose-Fermi theory will be
sensible. However, by making such an approximation we explicitly
neglect processes by which a zero of the wavefunction is
transferred from one electron to another (changing a boson into a
fermion or vice versa).

Since at $\nu=\half$ the effective magnetic field seen by bosons
or fermions is zero,  we will take $\Phi_f^\alpha$ to be a filled
Fermi sea of $N^\alpha_f$  CFs in layer $\alpha$. Similarly the
CBs in zero field condense into a $\vec k=0$ ground state (as
assumed in the wavefunction). If we have two layers of matched
density we must have $N^1_f = N^2_f$ in the ground state. However,
the overall number of CFs versus CBs is a matter of energetics and
presumably varies continuously as $d$ changes. When $N_f = 0$ in
both layers, this wavefunction is the 111 state, whereas $N_b = 0$
(or $N_f=N$) consists of 2 uncorrelated layers of CFs.

\begin{figure}[htbp]
\scalebox{.4}{\includegraphics*{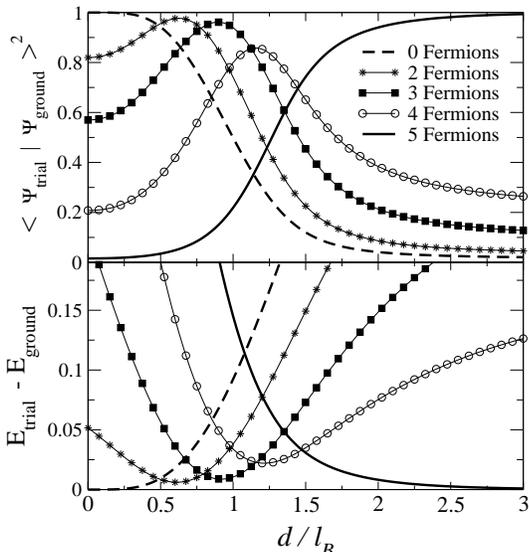}}  \caption{(Top)
Overlap squared of trial wavefunctions with exact ground state as
a function of layer spacing $d/\ell_B$. (Bottom) Energy of trial
states minus energy of the of ground state as a function of
$d/\ell_B$.  Calculations are done with 5 electrons per layer on a
bilayer sphere with flux 9$\phi_0$.  At low spacing the 111 state
($\state{0}$) has the highest overlap and the lowest energy.
However as the spacing increases, we go through a sequence of
states $\state{2}$, $\state{3}$, and $\state{4}$ and finally at
large spacing $\state{5}$ (the CF Fermi sea) has the highest
overlap and lowest energy.}
\end{figure}

{\it Numerical Calculations:}  We have considered a finite sized
bilayer sphere with 5 electrons per layer and a monopole of flux
$9 \phi_0$ at its center.   We first generate mixed CB-CF
wavefunctions of the form of Eq. \ref{eq:transitionwf}. The method
of numerical generation is involved and will be discussed in a
forthcoming paper. We note that in the Jastrow factors of Eq.
\ref{eq:transitionwf} the fermions ($z_i, w_i \leq N_f$)
experience one less flux quantum than the bosons ($z_i, w_i >
N_f$).  Thus, when the CBs are in zero effective magnetic field
(so they can condense), the CFs experience a single flux quantum.
Thus, $\Phi_f$ is modified to represent fermions on a sphere with
a monopole of charge $\phi_0$ in the center.  We generate
wavefunctions with 0,1,2,3,4,5 fermions per layer (and
correspondingly 5,4,3,2,1,0 bosons per layer). By construction the
1 fermion state is identical to the 0 fermion state. However all
of the remaining states are linearly independent. Thus we have
generated 5 states which we label $\state{0}$, $\state{2}$,
$\state{3}$, $\state{4}$, and $\state{5}$. Of course $\state{0}$
is the 111 state, and $\state{5}$ is two separate layers of
composite fermions in the presence of 1 flux quanta. (In most of
the wavefunctions, the fermionic ground state in a single
layer\cite{EdandNick} is at maximal angular momentum $L \neq 0$.
We always couple these to produce a bilayer $L=0$ wavefunction).

We next perform an exact lowest Landau level diagonalization on
the bilayer sphere using a pure coulomb interaction
$v_{11}(r)=e^2/r$ within a layer and $v_{12}(r) =e^2/\sqrt{r^2 +
d^2}$ between the two layers where $r$ is the chord distance
between two points. We vary the spacing $d$ between the two layers
and obtain exact ground states at each spacing $d$. In Figure 1,
we show the overlap (squared) of each of our trial wavefunctions
with the exact ground state at each of the values of $d/\ell_b$.
We also show the relative energy of the various ground states as a
function of $d/\ell_B$.  It is clear that the mixed CB-CF states
have very high overlap with the exact ground state and very low
energy in the transition region.  This supports the picture of the
transition from CF to CB occurring through a set of ground states
with interpenetrating CF-CB mixtures.

In Figure 2 we show the inter- and intra-layer electron pair
correlation functions ($g_{12}$ and $g_{11}$).  We see that the
mixed CB-CF states allow us to interpolate between the types of
correlations that exist in the limiting 111  and Fermi liquid
states.

{\em Chern-Simons RPA:} We can  calculate the resistivities and
drag resitivities of these mixed Bose-Fermi states by using the
Chern-Simons RPA approach.   Analogous to Eqs. \ref{eq:Beff} and
\ref{eq:Beff2} we can write effective electric (mean) fields  seen
by the bosons or fermions in  layer $\alpha$:
\begin{eqnarray} \label{eq:Eeff}
  {\cale}^{\alpha}_f &=& \vec E^\alpha - 2 \beps \vec J^\alpha
 - \beps (\vec J_b^1 + \vec J_b^2) \\
  {\cale}^{\alpha}_b &=& \vec E^\alpha -  \beps (\vec J_b^1
+ \vec J_b^2 + \vec J_f^1
  + \vec J_f^2) \label{eq:Eeff2}
\end{eqnarray}
where $\vec J_{f(b)}^\alpha$ is the Fermi (Bose) current in layer
$\alpha$ with the total current in layer $\alpha$ given by $\vec
J^\alpha =\vec J_b^\alpha + \vec J_f^\alpha$.

We supplement Eqs. \ref{eq:Eeff}, \ref{eq:Eeff2} with transport
equations for the fermions (bosons) in each layer
$\cale^\alpha_{f(b)} = \rho^\alpha_{f(b)} \vec J_{f(b)}^\alpha$.
Finally,  for a drag experiment we fix $\vec J^2=0$ and fix $\vec
J^1$ finite.  We then solve for $\vec E^1$ and $\vec E^2$ yielding
the in-layer resistivity ($\vec E^1 = \rho^{11}\vec J^1$) and the
drag resistivity ($\vec E^2 = \rho^D \vec J^1$).  Assuming layer
symmetry so $\rho_{b(f)}^\alpha$ is not a function of $\alpha$,
the results of such a calculation yield $\rho^{11} = (G + H)/2$
and $\rho^D = (G-H)/2$ where $G = (\rho_b^{-1} + \rho_f^{-1})^{-1}
+ 2 \beps$, and $H^{-1}=\rho_b^{-1} + (\rho_f + 2 \beps)^{-1}.$ At
filling fraction $\nu_1 = \nu_2 = \half$ both $\rho_b$ and
$\rho_f$ are diagonal (since CBs and CFs are in zero effective
field). Thus, there are only two free parameters ($\rho_{b,xx}$
and $\rho_{f,xx}$) and four measurable quantities ($\rho_{xx},
\rho_{xy}, \rho^D_{xx}, \rho^D_{xy}$) which enables us to derive
experimental predictions, such as $\rho^D_{xy} + \rho^{11}_{xy} =
4 \pi \hbar/e^2$\cite{end2}. Other more complicated
relationships can also be derived.

\begin{figure}[tbp]
\scalebox{.35}{\includegraphics*{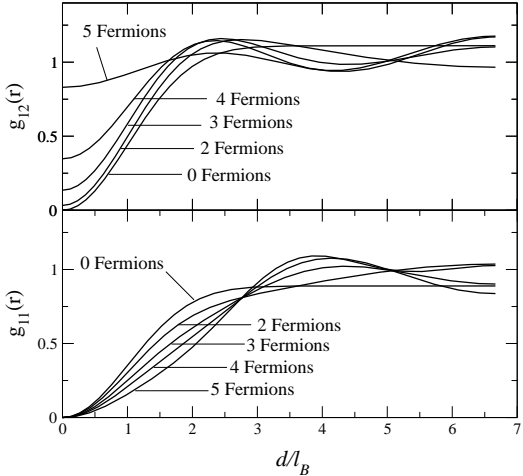}}
  \caption{Interlayer(top) and Intralayer (bottom) electron pair correlation
  functions $g_{12}$ and $g_{11}$ as a function of (arc) distance for each of
   the trial wavefunctions
  (5 electrons per layer again). As we go from the Fermi liquid
  state($\state{5}$)
  to the 111 state ($\state{0}$), replacing  CFs with CBs, the short range
  intralayer correlations are reduced and the interlayer
  correlations build up.  The deviation of $g_{12}$ from unity in the Fermi liquid
  state  is caused by the fact that two $L \neq 0$
  single layer states are combined to form an $L=0$ bilayer state.  }
\end{figure}

When $d$ is large, we assume that there are very few CBs and thus
$\rho_{b,xx}$ should be large (at least at finite temperature or
with disorder). Furthermore, from the experimentally measured CF
resistivity, it is clear that $\rho_{f,xx}$ is small ($\ll 2 \pi
\hbar/e^2$) at least at large $d$. As $d$ is reduced, presumably
the density of CBs increases and $\rho_{b,xx}$ drops until the CBs
condense at some critical density and $\rho_{b,xx}=0$ (yielding
perfect Hall drag as in the 111 state ). Simultaneously, as $d$
decreases, $\rho_{f,xx}$ presumably increases, but only
mildly\cite{HLR} and may remain small even when the CBs condense.
Only when the density of CFs is near zero should $\rho_{f,xx}$
diverge.  In the limit that $\rho_{f,xx} \ll 2 \pi \hbar/e^2$ it
is easy to derive the above mentioned semicircle law
$(\rho^D_{xx})^2 + (\rho^D_{xy} + \pi \hbar /e^2)^2 \approx ( \pi
\hbar /e^2)^2 $ from our above expression for $\rho^D$. In
addition, in this limit one can derive $\rho^{11}_{xx} \approx
\rho^D_{xx}$ which is also reasonably consistent with published
data\cite{Kellogg2}.

Our mixed CB-CF  theory can be generalized for unequal densities
as well as for filling fractions away from $\nu_1 = \nu_2 =
\half$.  Further, one may be able to treat the effects of an
in-plane magnetic field. Indeed, such an approach was already used
in Ref. \onlinecite{Milica} to understand bilayer tunnelling
experiments in tilted fields.

In summary, we have constructed a theory describing the crossover
between a CF liquid at large $d$ to the 111 CB state at small $d$
which can be thought of as a two-fluid model.  Comparisons of
trial wavefunctions to exact diagonalization are very favorable,
and the corresponding Chern-Simons transport theory appears to be
in reasonable agreement with experimental data.

Acknowledgements: The authors would like to acknowledge helpful
conversations with J. Eisenstein and B. I. Halperin.  MM
acknowledges support from Grant number 1899 of the
 Serbian Ministry of Science, Technology, and Development.
 EHR acknowledges support from DOE under contract DE-FG03-02ER45981 .

\end{multicols}
\end{document}